\documentclass[reviewcopy]{elsart}
\usepackage[nomarkers]{endfloat}

\usepackage{amstext}
\usepackage{amsmath}
\usepackage[dvipdf]{graphics}
\usepackage{epsfig}
\usepackage{pslatex}

\journal{IFES 2004}

\begin{document}
\begin{frontmatter}
\title{An improved reconstruction procedure for the correction of local magnification effects in {3DAP}}
\author[rouen,crv]{F. De Geuser}
\author[rouen]{W. Lefebvre\corauthref{corresp}}
\author[rouen]{F. Danoix}
\author[rouen]{F. Vurpillot}
\author[sintef]{B. Forbord}
\author[rouen]{D. Blavette}
\address[rouen]{Groupe de Physique des Mat{\'e}riaux -- UMR CNRS 6634,
Institut des Mat{\'e}riaux de Rouen,
76801 Saint-Etienne-du-Rouvray Cedex -- France}
\address[crv]{Pechiney CRV (Alcan Group) --  
BP 27, 38241 Voreppe Cedex -- France}
\address[sintef]{Norwegian University of Science and Technology -- 7491 Trondheim, 
and SINTEF Materials Technology -- 7465 Trondheim -- Norway}
\corauth[corresp]{Corresponding author. Address: GPM -- UMR CNRS 6634,
Institut des Mat{\'e}riaux de Rouen -- BP 12,
76801 Saint-Etienne-du-Rouvray Cedex -- France. Tel.: +33 232 95 51 41; Fax.: +33 232 95 50 32; E-mail: williams.lefebvre@univ-rouen.fr}

\begin{abstract}
A new 3DAP reconstruction procedure is proposed, which accounts for the evaporation field of a secondary phase. It applies the existing cluster selection softwares to identify the atoms of the second phase and subsequently an iterative algorithm to homogenise the volume laterally. This procedure, easily implementable on existing reconstruction software, has been applied successfully on simulated and real 3DAP analyses.
\end{abstract}
\begin{keyword}
Atom probe, local magnification, data analysis
\end{keyword}

\end{frontmatter}

\section{Introduction}

One of the major problems that one has to face to with 3DAP originates from the difference of evaporation field between phases. This commonly happens when studying precipitation for instance. The difference of evaporation behaviour can provoke severe atomic density variations in the 3D reconstructed volume and biased estimations of the morphologies of the observed objects. Although corrections of the data in 1DAP have been proposed (\cite{miller91}), 3DAP data correction are difficult. Simulations of ion trajectories (\cite{vurpillot00}) have shown that two different effects occur: trajectories overlaps at the interface and the so called local magnification effect.

A new reconstruction procedure for correcting this latter effect is proposed. It can be proceeded by adjusting the reconstruction parameters (i.e. evaporation field) whether a matrix atom or a precipitate atom is considered. Combining this correction to an algorithm that corrects the coordinates in the plane perpendicular to the analysis is the proposed procedure to account for the difference in magnifications between the matrix and the second phase in the reconstruction protocol.

\section{Local magnification effect in 3D atom probe analysis reconstruction}
\label{sec:recons}

The basic principles of 3DAP are provided elsewhere \cite{miller}. The volume reconstruction methods are based on the fact that the atoms are evaporated atomic layer by atomic layer. Simple geometric considerations are often enough to resolve atomic planes, proving the efficiency of this simple method \cite{bas95}. In a first approximation, a 3D atom probe can be considered as a point projection microscope. The sample is prepared as a tip that can be characterised by its radius of curvature $R$. This radius of curvature determines the magnification $\eta$ of the projection. $R$ is related to the total voltage $V$ applied to the specimen by the relation: 
\begin{equation}
	R = \frac{V}{F\beta} \label{eq:radius}
\end{equation}
where $F$ is the evaporation field and $\beta$ a geometrical factor. According to this relation, the magnification of the projection can be computed for each evaporated atom and its initial position $(x,y)$ at the specimen surface can be deduced from the position of the impact on the detector. The third coordinate $z$ is incremented for each atom by the ratio between its atomic volume $\Omega$ and the analysis surface $S_a$. $S_a$ is deduced from the detector surface $S_d$ and from the magnification $\eta$.

In order to study the effect of the reconstruction procedure on basic systems, we have simulated the field evaporation of an $A-B$ alloy containing pure $B$ spherical precipitates embedded in a pure $A$ matrix. The precipitates are coherent with the matrix, and with an evaporation field 1.25 times larger than that of the matrix. The atoms are detected on a virtual detector with a given surface. For each detected atom, the position on the detector and the initial position in the volume were evaluated.

\begin{figure}
	\begin{center}
		\includegraphics[width=8cm]{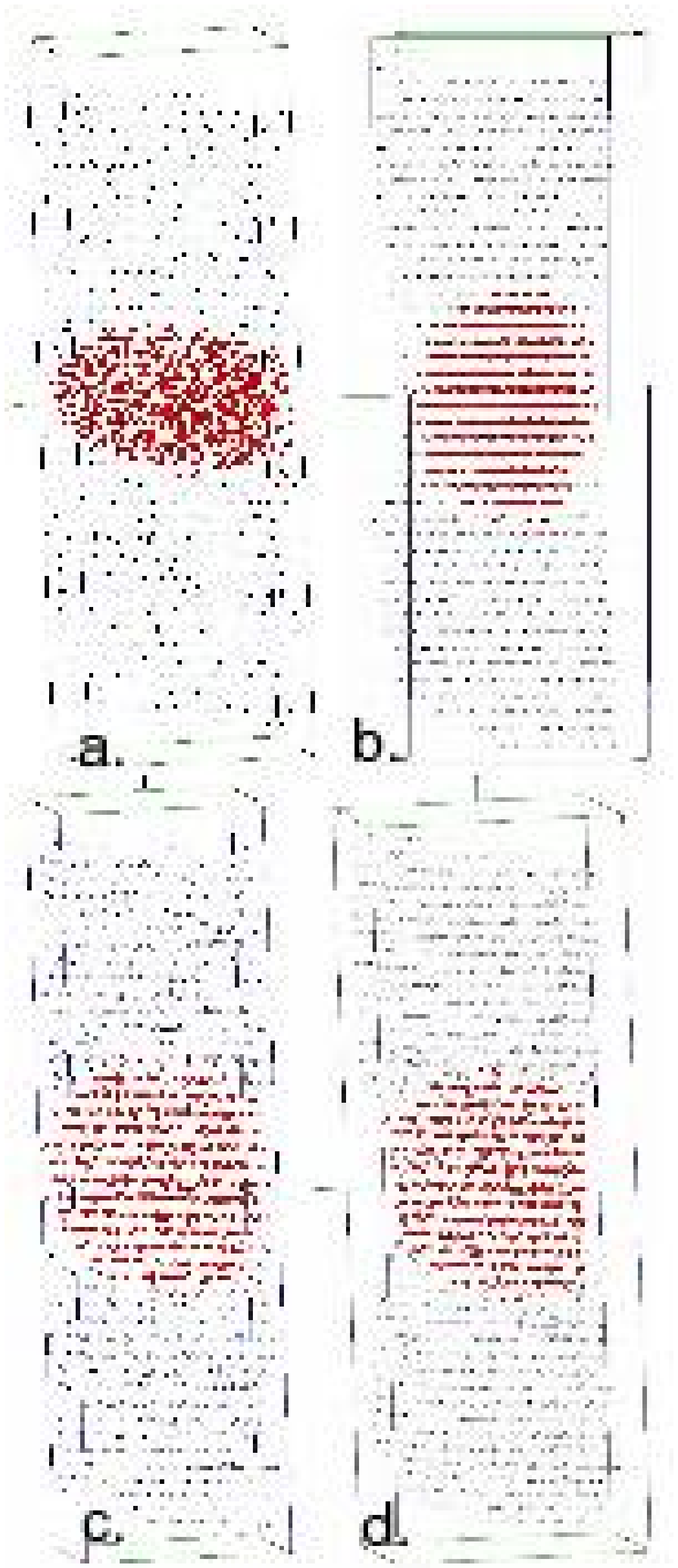}
		\end{center}
			\caption{\label{fig:simul}Simulation of a 3DAP analysis on an A-B alloy with a spherical precipitate. The relative evaporation field of the second phase is 1.25. a. Standard reconstruction method. b. Initial positions of the atoms c. Correction in depth d. Correction in depth and in the $(x,y)$ plane.}
		\end{figure}

The volume reconstructed by the standard method is shown on Fig. \ref{fig:simul}a while Fig. \ref{fig:simul}b represents the positions of the detected atoms in the initial structure. The reconstructed precipitate in Fig. \ref{fig:simul}a exhibits a lens shape and it is difficult to resolve any atomic planes (they can be seen but are very close to each other). Strong atomic density variations appear in the precipitate and at its interface with the matrix. This effect originates from the difference in evaporation field between the two phases. In order to evaporate the precipitate, the tip  develops locally a different radius of curvature according to Eq. \ref{eq:radius}, and the magnification changes accordingly.

It has been proposed in the past to adjust the position of the atoms to achieve an equal density in the volume (\cite{thesefrancois}). The method was based on a relaxation method. Each atom applies on its neighbours a force derived from a potential (e.g. harmonic or Lennard-Jones type). The system is then solved numerically until equilibrium is reached. One of the hypothesis is that the volume limits remain fixed. However, as the atoms are moved isotropically until they are all equally spaced, the resolution of atomic planes is necessarily lost. This leads to a kind of amorphization of the volume. Furthermore, the fixed volume limits conditions is too strict to be applied to large precipitates where the variation in the analysis surface is the determinant factor. This is clearly demonstrated in Fig. \ref{fig:simul}b.

A possible way to overcome the local magnification effect is to apply the appropriate reconstruction parameters for the matrix and for the precipitates. A basic requirement is to be able to select the atom of the second phase. For that purpose, it is possible to run cluster selection type softwares (\cite{heinrich03,vaumousse03}) despite the local magnification effect. An alternative option is to use the density variation for the cluster identification using Fourier analysis \cite{vurpillot04}. Once the selection is achieved, it becomes  possible to run the reconstruction algorithm, taking the second evaporation field into account.

Although this simple correction can be applied successfully to the coordinate along the analysis direction (see Fig. \ref{fig:simul}c), the lateral coordinates correction is not straightforward because of the interdependence between the atoms coordinates (\cite{bas95}). To solve this problem, we use an iterative  algorithm which homogenizes the volume by superimposing a grid to the 3D atom map, and deforming this grid according to the gradient of atomic density.

The nodes of the grid are only slided in the $(x,y)$ planes, i.e. the dimension in the $z$ scale is not modified by this algorithm. Slice by slice, the atomic density is homogenised. The steps of the grid are larger than the lattice parameter, typically between 1 and 2 nm, in order to keep the structural information at the lattice scale. When a grid becomes too distorted (the limit is fixed in terms of angles), the new atom  coordinates are computed and another grid is superimposed to the resulting volume. For each iteration, the mean quadratic difference between the density of each box and the expected density is checked. The program proceeds until convergence is achieved. An important advantage of this method is that no condition on the reconstructed volume is needed.

The final reconstructed volume is shown on Fig. \ref{fig:simul}d. The atomic planes are shown to be well resolved and properly spaced. The density is homogeneous. The shape is very similar to that of the initial positions of the atoms in the sample  (Fig. \ref{fig:simul}b), which proves the efficiency of the proposed protocol. In particular, the variation of the analysed surface is properly accounted for and corrected. This procedure will now be applied to various systems where the difference in evaporation field between matrix and precipitates is important.

\section{Application to low field precipitates}

The algorithm described in section \ref{sec:recons} has been applied on two typical systems where the evaporation field of the precipitates is lower than that of the matrix, namely the FeCu and the AlZnMg system. 

\begin{figure}
	\begin{center}
		\includegraphics[width=13cm]{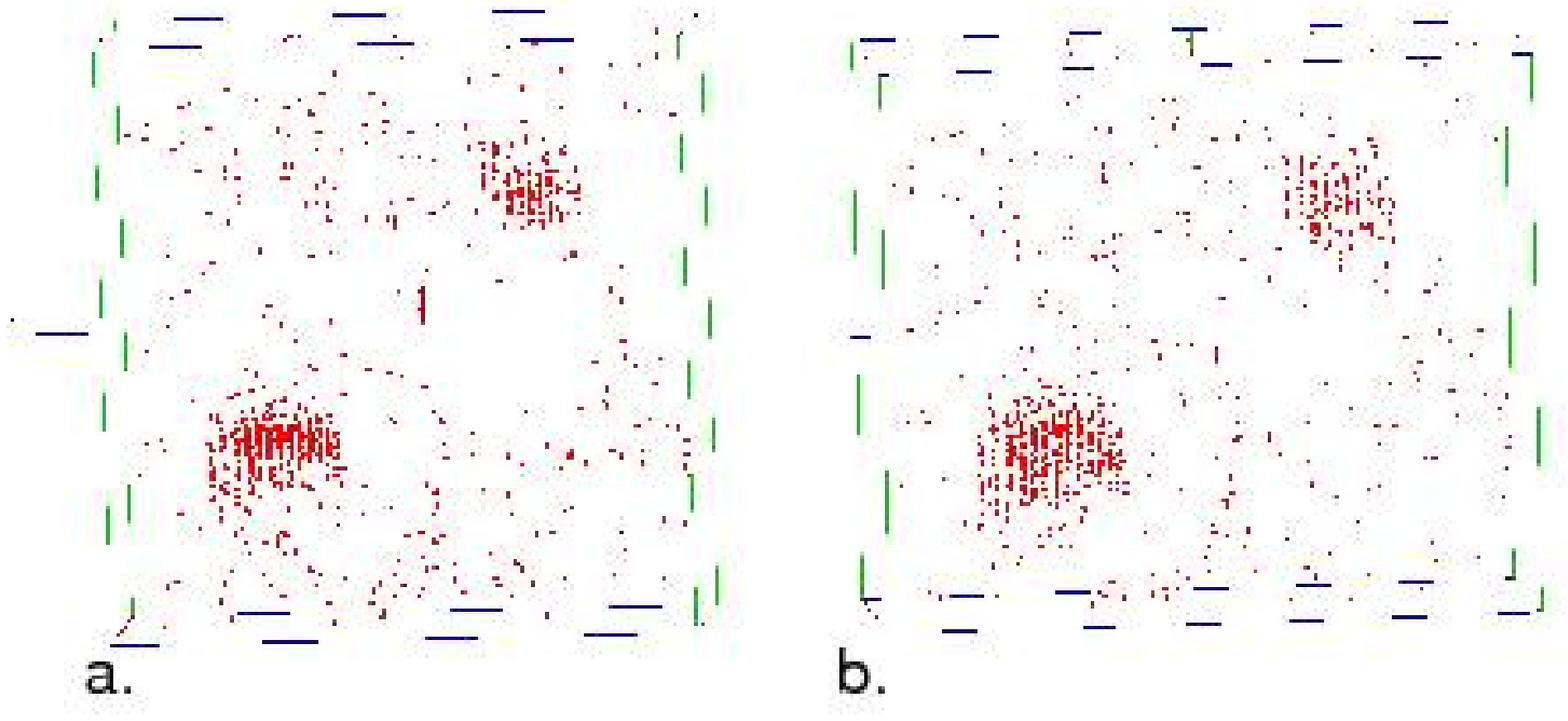}		
		\end{center}
			\caption{\label{fig:fecu}3D reconstruction of a copper particle in a $FeCu$ alloy: a. Standard protocol b. Correction in depth and in the $(x,y)$ plane. The elongation of the particle disappear but the coherent planes remain.}
		\end{figure}

The FeCu specimen studied contains spheroidal pure copper precipitates that are coherent with the iron matrix (\cite{pareige95}). When standard reconstruction procedures are used, the lower evaporation field of the Cu precipitates leads to ellipsoidal shapes of the precipitates, elongated along the analysis direction (Fig. \ref{fig:fecu}a). Applying the improved procedure, the shape of the precipitates is more spheroidal, as expected (Fig. \ref{fig:fecu}b). Atomic planes in the copper precipitates can be observed before and after correction.

			\begin{figure}
			\begin{center}
		\includegraphics[width=8cm]{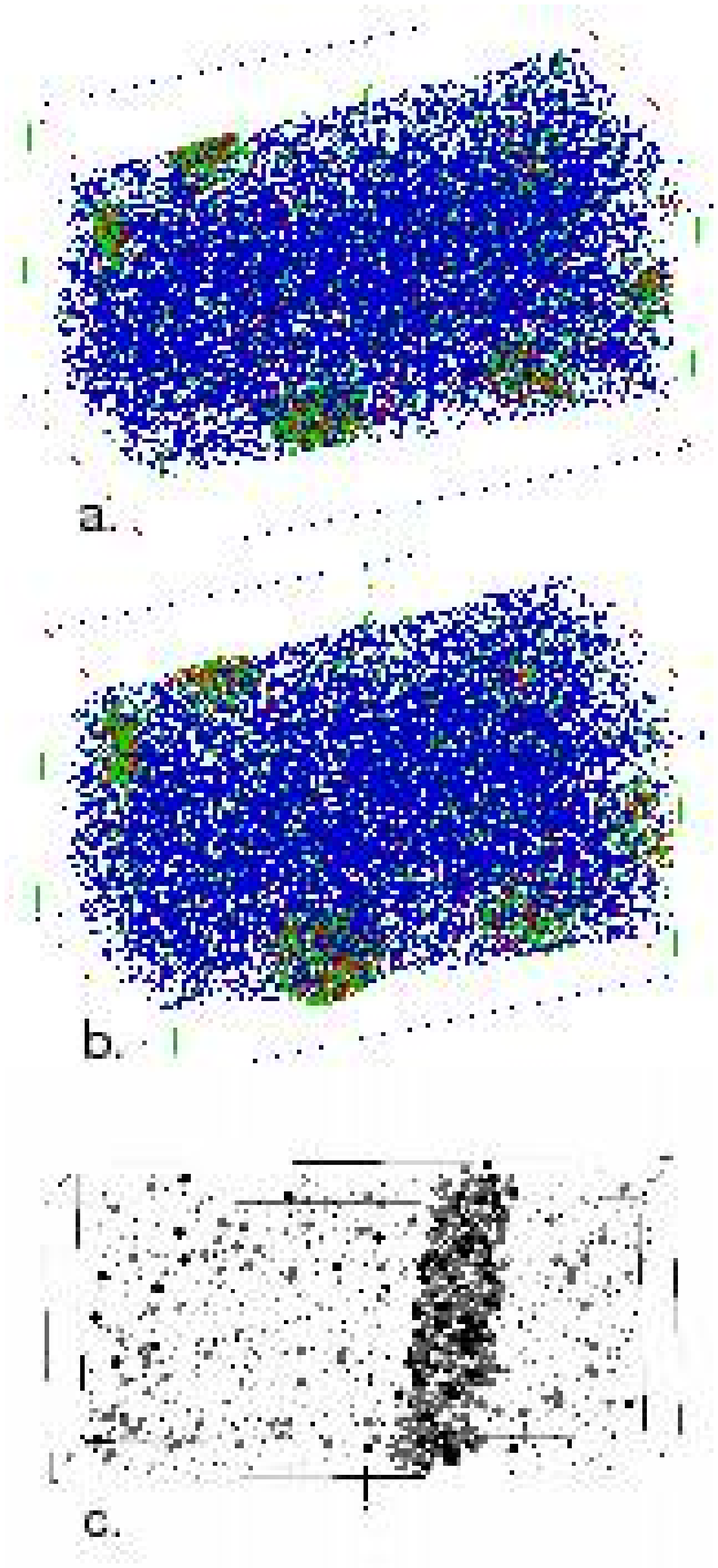}
		\end{center}
			\caption{\label{fig:7xxx}3D reconstruction of a $AlZnMg$ alloy containing Zn- and Mg-rich particles : a. Standard protocol b. Correction in depth and in the $(x,y)$ plane c. $\eta'$ platelet parallel to the $Al$ \{111\} planes.}
		\end{figure}	

Another illustration is given from an AlZnMg alloy, which contains $\eta'$ platelets  parallel to the \{111\} planes of the matrix and $\eta$ precipitates that are more or less spherical (\cite{deschamps01}). Both $\eta$ and $\eta'$ have a lower field of evaporation with respect to the matrix. The number density of atoms in the precipitates is 3 times larger than in the matrix. After application of the improved  protocol, their morphologies are modified (Fig. \ref{fig:7xxx}b). The atomic density becomes homogeneous. Fig. \ref{fig:7xxx}c is a zoom on the $\eta'$ platelet that can be seen at the top left corner of the volume on Fig. \ref{fig:7xxx}b. This platelet is shown to be still properly aligned on the \{111\} planes, which indicates that the reconstruction is valid even at the atomic scale.

\section{Application to high field precipitates}
\label{highfield}

\begin{figure}
\begin{center}
		\includegraphics[width=8cm]{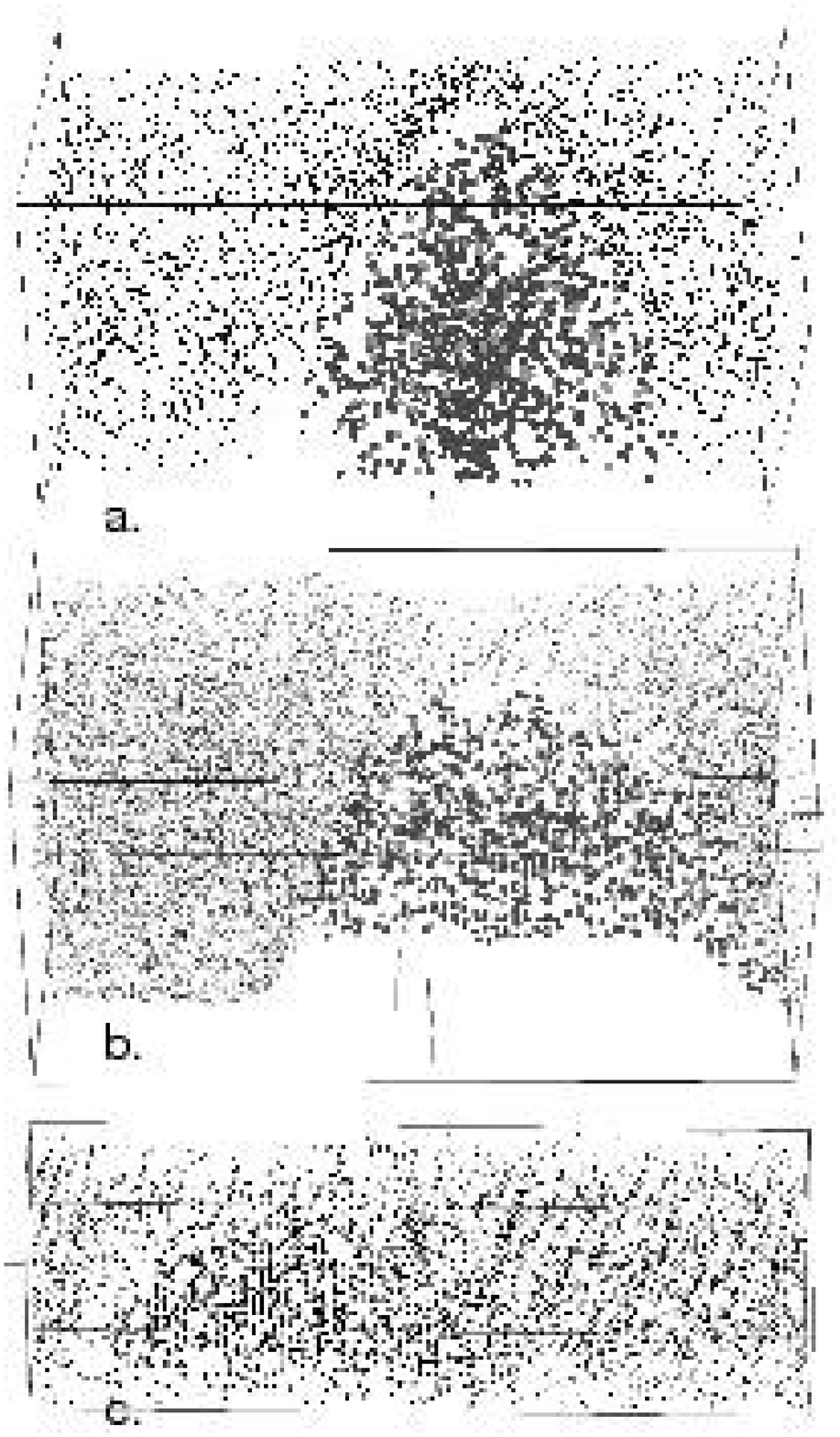}
		\end{center}
			
			\caption{\label{fig:alzrsc}	\label{fig:alzrsc1}	\label{fig:alzrsc2}	\label{fig:alzrsc3}3D reconstruction of a $Al_3(Zr,Sc)$ particle following : a. Standard protocol b. Correction in depth and in the $(x,y)$ plane c. \{110\} planes of the particle and \{220\} planes of the matrix simultaneously resolved.}
		\end{figure}

We have applied the method to an AlZrSc alloy. The microstructure of the alloy consists in $L1_2$ $Al_3(Zr,Sc)$ spherical particle in a $FCC$ aluminium matrix. These dispersoids have a much higher evaporation field than the matrix \cite{forbord04}. In the analysis conditions used, their size is in the same order of magnitude as the analysis surface. Because of this, when they are intercepted, the standard reconstruction procedure compensates the low density in the precipitate by squeezing the volume in depth. As the area ratio between matrix and precipitate changes during the interception of the precipitate, this effect is continuous. This can be seen on figure \ref{fig:alzrsc}a where the matrix is subjected to a strong density variation in the vicinity of the precipitate. The  \{220\} planes of the matrix are resolved and the interplanar spacing is coherent with the lattice parameter which indicates that the reconstruction parameter combination is correct for this phase. Next to the precipitate, however, these atomic planes are closer and quite blurred. A former treatment (\cite{forbord04}) has shown that an adjustment of the reconstruction parameters is necessary to reveal the \{110\} superstructure planes in the precipitate. It was difficult, though, to resolve the superstructure sufficiently in the entire precipitate. This was due to the continuous variation of the surface ratio on the detector between the two phases. Furthermore, with this adjustment, it was not possible to image simultaneously the atomic planes in the matrix.

The application of the proposed method shows simultaneously the \{220\} planes in the matrix and the \{110\} planes in the particle, with a distance of 0.145 nm and 0.29 nm respectively (see Fig \ref{fig:alzrsc3}c), in excellent agreement with the lattice parameter. The spherical morphology is also shown to be better reconstructed. No remaining atomic density variation can be observed.

\section{Discussions}
\label{sec:Discussions}

The proposed reconstruction procedure obviously improves the 3D reconstruction of precipitates embedded in a matrix. With this method, the particles morphology can be resolved without any preliminary assumption. The local structural information contained in the analysis, such as atomic planes resolution for instance, is not affected by the improved procedure. On the contrary, it can resolve details that were invisible on the standard reconstruction volumes.

Furthermore, it takes into account the anisotropy of the local magnification effect considered along or perpendicularly to the analysis direction. In the standard reconstruction method, this anisotropy will lead to biases in the results from data analysis, such as concentration profiles or proxigrams (\cite{hellman02}), that consider all the precipitates in the same concentration profile. Indeed, two precipitates of the same dimensions but intercepted differently by the surface of analysis are subject to distinct effects because of their different interception area.

\begin{figure}
	\begin{center}
		\includegraphics[width=11cm]{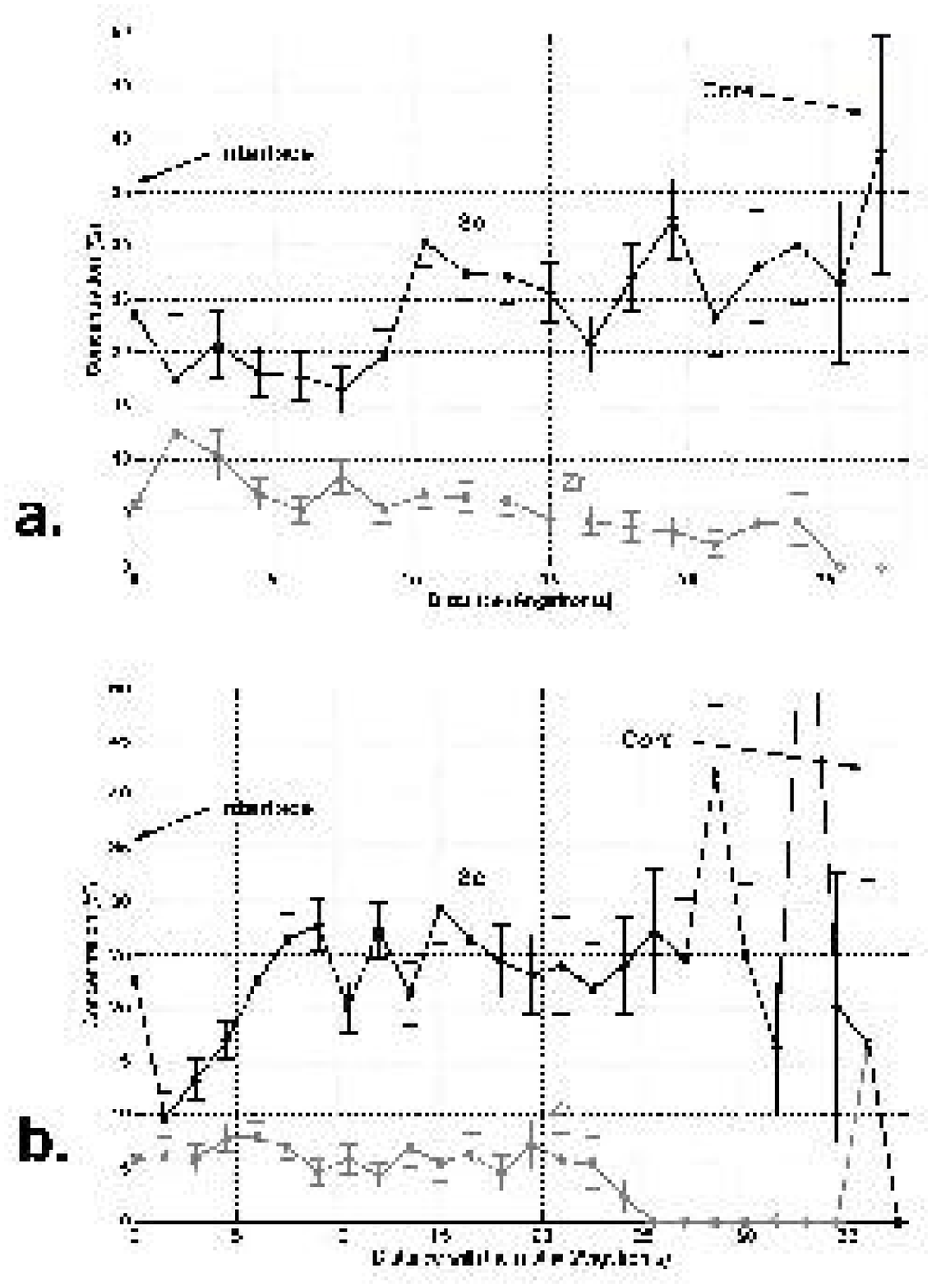}
		\end{center}
			\caption{\label{fig:erosion}Erosion concentration profile for the zirconium on the $Al_3(Zr,Sc)$ particle from Fig. \ref{fig:alzrsc}. a. Uncorrected volume b. Corrected volume, revealing a Zr-poor core.}
		\end{figure}

But even on independently treated precipitates, the bias on the geometry is not negligible. We have plotted an erosion profile of the first particle. This profile (Fig. \ref{fig:erosion}) corresponds to a binning of the atoms of the precipitates as a function of their distance to the matrix. The concentration is plotted as a function of the distance. The bins step is 0.15 nm. The error bars are the statistical errors on the concentration due to sampling (related to the number of atoms in the bin). Fig. \ref{fig:erosion}a is the profile with the "standard" volume and Fig. \ref{fig:erosion}b with the "corrected" volume. The expected Zr--poor core \cite{forbord04,theseclouet} of the particle can be clearly indicated on Fig. \ref{fig:erosion}b, whereas on Fig. \ref{fig:erosion}a the bins overlap and hence average it out. In addition, the Zr concentration is continuously decreasing on Fig. \ref{fig:erosion}a, instead of forming a plateau, as in Fig. \ref{fig:erosion}b.

\section{Conclusion}

A new reconstruction protocol has been proposed. It uses existing cluster selection softwares on a primary reconstructed volume to attribute another evaporation field to the second phase. After  correction of the analysed depth, the atomic density in the $(x,y)$ section is homogenised by an iterative algorithm. The resulting volumes are shown to be closer to reality than the standard reconstruction. This is confirmed by the performed simulation. Primary unresolved details can be seen on reconstructed volume, such as atomic planes both in precipitates and matrix. Concentration gradients in particles are shown to be more accurate. This method can be easily implemented on an existing reconstruction algorithm in order to be routinely used.


\end{document}